\documentclass[reprint,amsmath,amssymb,aps,prl,superscriptaddress,longbiography]{revtex4-2}

\usepackage{graphicx}
\usepackage{dcolumn}
\usepackage{bm}
\usepackage{paralist}
\usepackage{color}
\usepackage{url}
\usepackage{tikz}

\definecolor{darkyellow}{rgb}{1,0.84,0}
\def\d{\mathrm{d}}

\begin{document}

\title{Unified Flow Rule of Undeveloped and Fully Developed Dense Granular Flows Down Rough Inclines}

\author{Yanbin Wu}
\author{Thomas P\"ahtz}
\email{0012136@zju.edu.cn}
\author{Zixiao Guo}
\affiliation{Institute of Port, Coastal and Offshore Engineering, Ocean College, Zhejiang University, Zhoushan, China}
\author{Lu Jing}
\affiliation{Institute for Ocean Engineering, Shenzhen International Graduate School, Tsinghua University, Shenzhen, China}
\affiliation{State Key Laboratory of Hydroscience and Engineering, Tsinghua University, Beijing, China}
\author{Zhao Duan}
\affiliation{College of Geology and Environment, Xi'an University of Science and Technology, Xi'an, China}
\affiliation{Shaanxi Provincial Key Laboratory of Geological Support for Coal Green Exploitation, Xi'an, China}
\author{Zhiguo He}
\email{hezhiguo@zju.edu.cn}
\affiliation{Institute of Port, Coastal and Offshore Engineering, Ocean College, Zhejiang University, Zhoushan, China}

\date{\today}

\begin{abstract}
We report on chute measurements of the free-surface velocity $v$ in dense flows of spheres and diverse sands and spheres-sand mixtures down rough inclines. These and previous measurements are inconsistent with standard flow rules, in which the Froude number $v/\sqrt{gh}$ scales linearly with $h/h_s$ or $(\tan\theta/\mu_r)^2h/h_s$, where $\mu_r$ is the dynamic friction coefficient, $h$ the flow thickness, and $h_s(\theta)$ its smallest value that permits a steady, uniform dense flow state at a given inclination angle $\theta$. This is because the characteristic length $L$ a flow needs to fully develop can exceed the chute or travel length $l$ and because neither rule is universal for fully developed flows across granular materials. We use a dimensional analysis motivated by a recent unification of sediment transport to derive a flow rule that solves both problems in accordance with our and previous measurements: $v=v_\infty[1-\exp(-l/L)]^{1/2}$, with $v_\infty\propto\mu_r^{3/2}\left[(\tan\theta-\mu_r)h\right]^{4/3}$ and $L\propto\mu_r^3\left[(\tan\theta-\mu_r)h\right]^{5/3}h$.
\end{abstract}

\maketitle

Dry, dense granular flows down rough inclines have been responsible for numerous natural disasters in recorded history, including one of the worst known to mankind: In 1970, the Ancash earthquake in Peru triggered the Huascar\'an rock and ice avalanche, which killed around 25,000 people in the process \citep{KeeferLarsen07,SchusterHighland07,Petley12}. The arguably simplest laboratory system one can study to gain physical insights into such destructive natural flows is a homogeneous chute flow (Fig.~\ref{Sketch}):
\begin{figure}[t]
 \includegraphics[width=246pt]{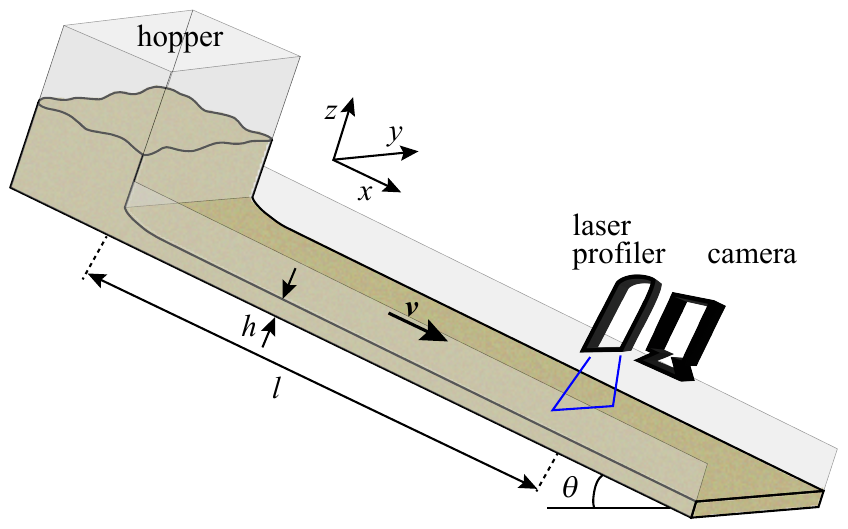}%
 \caption{Sketch of chute flow experimental setup.}
 \label{Sketch}
\end{figure}
a controllable amount of nearly monodisperse granular material is released and subsequently accelerated down a rough incline until it reaches a fully developed or steady flow state. In principle, this design permits measuring the spatial evolution of the flow velocity $v$. However, to our knowledge, experimental studies based on this~\citep{Pouliquen99,Ancey01,BorzsonyiEcke07} or similar~\citep{Campbelletal85,SadjadpourCampbell99,LougeKeast01,Goujonetal03,FelixThomas04,Deboeufetal06,Lougeetal15} designs have exclusively focused on determining the flow velocity in the (sometimes only seemingly) fully developed state, $v_\infty$, ignoring proxies for the undeveloped state, such as the characteristic distance $L$ that $v$ needs to reach $v_\infty$. The latter may be similarly important for estimating the flow energy of natural avalanches and therefore their destructiveness. More importantly, as we will see, for the typically studied upper range of the flow thickness $h$, the chute length $l$ can be on the order of $L$, too short for flows to fully develop, and this has hindered the quest of finding a universal law governing $v_\infty$. In fact, the two existing measurements-based propositions for this so-called flow rule are somewhat disagreeing with one another and also known to not be universal across granular materials~\citep{Pouliquen99,BorzsonyiEcke07}:
\begin{align}
 v_\infty\sqrt{\cos\theta/h}&\propto h/h_s, \label{Scaling1} \\
 v_\infty\sqrt{\cos\theta/h}&\propto(\tan\theta/\mu_r)^2h/h_s, \label{Scaling2}
\end{align}
where $\mu_r$ is the dynamic friction coefficient, often measured as the tangent of the dynamic angle of repose~\citep{BorzsonyiEcke07}, and the stop thickness $h_s=h_s(\theta)$ the smallest value of $h$ that permits a steady, uniform dense flow state at a given inclination angle $\theta$. We nondimensionalize physical quantities using the grain density $\rho_p$, median grain size $d$, and base-normal gravitational acceleration $g\cos\theta$, implying that $\sqrt{gd\cos\theta}$ is the velocity and $\rho_pgd\cos\theta$ the stress scale.

\begin{figure}[ht]
 \includegraphics[width=246pt]{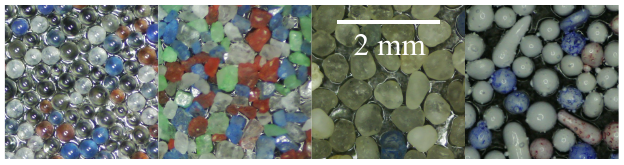}%
 \caption{Images of tested granular materials. From left to right: Spheres, Sand~1, Sand~2, Zirconia (Table~\ref{GranularMaterials}).}
 \label{GranularImages}
\end{figure}

In this Letter, we report experiments on dense granular flows in a laboratory chute using a range of granular materials: quartz spheres, two distinct natural sands, three spheres-sand mixtures of different mass proportions (3:1, 1:1, 1:3), and a zirconia (ZrO$_2$)-based mixture consisting of spheres and a few elongated grains ($14\%$ number fraction counted under the microscope) as a result of an imperfect manufacturing process (Fig.~\ref{GranularImages}, Table~\ref{GranularMaterials}).
\begin{table}[ht]
 \begin{tabular}{l|c|c|c|c|c}
  Granular material & Symbol & $d/\mathrm{mm}$ & $\mu_r$ & $\mu_1$ & $\mu_2$ \\ 
  \hline
  Spheres & \begin{tikzpicture}\filldraw[blue](0,0)circle(2pt){};\end{tikzpicture} & 0.35 & 0.35 & 0.36 & 0.77\\
  Sand~1 & \begin{tikzpicture}\filldraw[blue](0,0)rectangle(4pt,4pt){};\end{tikzpicture} & 0.32 & 0.60 & 0.58 & 1.35\\
  Sand~2 & \begin{tikzpicture}\filldraw[blue](0,-2pt)--(-2pt,0)--(0,2pt)--(2pt,0)--cycle{};\end{tikzpicture} & 0.50 & 0.54 & 0.55 & 1.53\\
	Spheres~BE~\citep{BorzsonyiEcke07} & \begin{tikzpicture}\draw[darkyellow,thick](0,0)circle(2pt){};\end{tikzpicture} & 0.36 & 0.41$\ast$ & 0.41 & 1.25\\
	Sand~BE~\citep{BorzsonyiEcke07} & \begin{tikzpicture}\draw[darkyellow,thick](0,0)rectangle(4pt,4pt){};\end{tikzpicture} & 0.40 & 0.59 & 0.60 & 2.03\\
	\hline
  Zirconia & \begin{tikzpicture}\filldraw[red](0,0)circle(2pt){};\end{tikzpicture} & 0.50 & 0.50 & 0.48 & 0.98\\
  3:1~Spheres-Sand~1 mixture & \begin{tikzpicture}\filldraw[red](0,0)rectangle(4pt,4pt){};\end{tikzpicture} & 0.34 & 0.47 & 0.47 & 1.28\\
  1:1~Spheres-Sand~1 mixture & \begin{tikzpicture}\filldraw[red](0,-2pt)--(-2pt,0)--(0,2pt)--(2pt,0)--cycle{};\end{tikzpicture} & 0.34 & 0.52 & 0.50 & 1.43\\
	1:3~Spheres-Sand~1 mixture & \begin{tikzpicture}\filldraw[red](-2pt,-1.1547pt)--(2pt,-1.1547pt)--(0,2.3094pt)--cycle{};\end{tikzpicture} & 0.33 & 0.57 & 0.54 & 1.42\\
 \end{tabular}
 \caption{Granular materials used in our and previous~\citep{BorzsonyiEcke07} experiments and their friction parameters ($\mu_1$ and $\mu_2$ as defined in Eq.~(\ref{hstop})). The angle of repose, and thus $\mu_r$, was measured through the tilting-box method as the angle of a heap formed by an avalanche~\citep{BeakawiAlHashemiBaghabraAlAmoudi18,SuppInclinedFlows}. ($\ast$) For Spheres~BE, $\mu_r$ was not reported and we therefore assume $\mu_r=\mu_1$.}
 \label{GranularMaterials}
\end{table}
Since we measure $v$ as the free-surface flow velocity, we combine our data with the only other existing data determined in the same manner: the sphere and natural-sand data of Ref.~\citep{BorzsonyiEcke07}. The combined data cannot be satisfactorily described by either Eq.~(\ref{Scaling1}) or Eq.~(\ref{Scaling2}). To overcome their lack of universality, we use simple arguments to derive an expression for the flow velocity in fully undeveloped conditions (i.e., the initial stage of flow development), $v_0$, and merge it with an expression for $v_\infty$ that does not depend on $h_s$. The latter is derived from a dimensional analysis that assumes that $v_\infty$ is primarily controlled by $\mu_r$ and the excess shear stress the sliding mass applies on the base. The resulting flow rule for $v$ incorporates a prediction for $L$ and agrees with the entire combined data across undeveloped and fully developed flow states.

\begin{figure}[ht]
 \includegraphics[width=246pt]{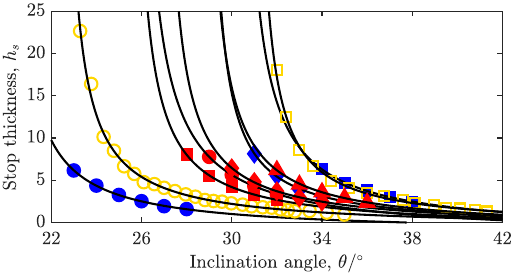}%
 \caption{Stop thickness curves $h_s(\theta)$ for the tested granular materials. The lines correspond to Eq.~(\ref{hstop}), with $\mu_1$ and $\mu_2$ as given in Table~\ref{GranularMaterials}. Note that values as low as $h_s\approx1$ can be accurately determined using our experimental method~\citep{BorzsonyiEcke07}.}
 \label{HstopCurve}
\end{figure}

The experiments are carried out in a $(2\times0.2)~\mathrm{m}$ chute (Fig.~\ref{Sketch}) with smooth sidewalls made of plexiglass and a base created by gluing spherical glass beads in an irregular manner onto a rubber surface. Their diameters vary between $0.6~\mathrm{mm}$ and $0.8~\mathrm{mm}$, about twice as large as the grain sizes of the tested materials, forcing a no-slip condition at the base~\citep{Jingetal16}. In an experimental run, a hopper feeds the material to the chute through an opening gate. Its height can be adjusted to control the flow thickness $h$, which rapidly approaches a quasisteady value, like in previous studies~\citep{Pouliquen99,BorzsonyiEcke06,BorzsonyiEcke07}. A laser profiler and high-speed camera record $h$ and the surface velocity $v$ from above the chute near the end of the flume, $l=1.7~\mathrm{m}$ away from the gate (for Zirconia, $l=1.3~\mathrm{m}$; in Ref.~\citep{BorzsonyiEcke07}, $l=1.55~\mathrm{m}$). This direct method is more accurate than measuring $v$ from the side or indirectly as the mean flow velocity from the mass flow rate~\citep{BorzsonyiEcke07}. Furthermore, in order to determine $h_s(\theta)$, we release material through (nearly) the smallest possible opening gate that allows for the development of a continuous flow across the entire chute at the given inclination angle $\theta$. Then, we close the gate, wait until the flow completely subsides, and determine $h_s=M/(\rho_pA\phi)$ from the mass $M$, packing fraction $\phi$, and area $A$ of the remaining deposit. This method, which was also used in Ref.~\citep{BorzsonyiEcke07}, yields an $h_s(\theta)$ curve of a functional form consistent with previous studies~\citep{Pouliquen99,BorzsonyiEcke07} (Fig.~\ref{HstopCurve}):
\begin{equation}
 h_s\propto(\mu_2-\tan\theta)/(\tan\theta-\mu_1), \label{hstop}
\end{equation}
where $\mu_1\approx\mu_r$ and $\mu_2$ are empirical coefficients obtained from fitting the measured curve to Eq.~(\ref{hstop}).

\begin{figure*}[t]
 \includegraphics[width=512pt]{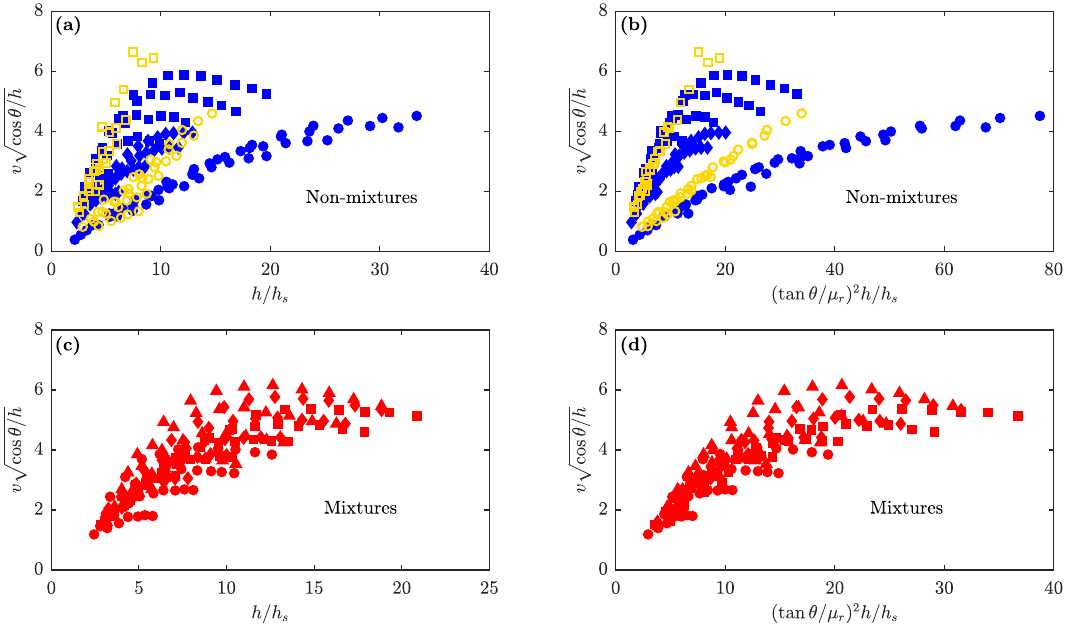}%
 \caption{Test of previous flow rules Eqs.~(\ref{Scaling1}) and (\ref{Scaling2}) against the measurements for ((a) and (b)) nonmixtures and ((c) and (d)) mixtures (Table~\ref{GranularMaterials}). These rules are not universal across granular materials, and the predicted proportionality between $v\sqrt{\cos\theta/h}$ and either ((a) and (c)) $h/h_s$ or ((b) and (d)) $(\tan\theta/\mu_r)^2h/h_s$ breaks down at sufficiently large $h/h_s$, where the curves tend to plateau (at different values for different inclination angles $\theta$).}
 \label{FlowRulesHstop}
\end{figure*}

We now test Eqs.~(\ref{Scaling1}) and (\ref{Scaling2}) against the experimental data using the measured values of $h_s$ (Fig.~\ref{FlowRulesHstop}).
It can be seen that Eq.~(\ref{Scaling2}) performs better than Eq.~(\ref{Scaling1}) and tends to agree with the data of a given granular material provided that $(\tan\theta/\mu_r)^2h/h_s$ is sufficiently low (Fig.~\ref{FlowRulesHstop}(b) and \ref{FlowRulesHstop}(d)). However, the slopes of the curves for different granular materials can deviate substantially from one another (Fig.~\ref{FlowRulesHstop}(b)). Furthermore, for larger values of $(\tan\theta/\mu_r)^2h/h_s$, the predicted proportionality between $v\sqrt{\cos\theta/h}$ and $(\tan\theta/\mu_r)^2h/h_s$ breaks down (also noticed in Ref.~\citep{BorzsonyiEcke07}) and the curves tend to plateau. In summary, neither of the standard flow rules is consistent with the measurements, both for nonmixtures and mixtures.

We hypothesize that the poor performances of Eqs.~(\ref{Scaling1}) and (\ref{Scaling2}) have mainly two reasons: First, these rules are not universal across granular materials. Second, in many of the measurements, including those of Ref.~\citep{BorzsonyiEcke07}, $v$ was not fully developed (i.e., $v<v_\infty$). This would explain why these rules tend to perform better for smaller values of $h/h_s$. In fact, previous discrete element method (DEM) simulations of inclined flows of spheres have demonstrated that the time $T$ a flow needs to fully develop increases with $h$~\citep{ParezAharonov15}. Below, we address both potential reasons separately.

To model undeveloped flows, we assume that $v$ follows a functional relationship of the form
\begin{equation}
 v/v_0=f(v_\infty/v_0), \label{MergeFunction}
\end{equation}
where $v_0=v_0(l)$ denotes the value of $v$ after traveling a distance $l$ for fully undeveloped conditions ($l\ll L$) and the function $f$ shall satisfy $f(X)\simeq X$ and $f(X)\simeq1$ for sufficiently small and large $X$, respectively. This form represents the simplest possible merging of $v=v_\infty$ with $v=v_0$ when $X$ increases. Furthermore, we assume that, for $l\ll L$, the flow obeys the same laws as an object sliding down an inclined plane, with $\mu_r$ playing the role of the sliding friction coefficient. That is, the flow is driven by the constant acceleration $a=g\sin\theta-\mu_rg\cos\theta$ and exhibits a velocity $v_0\propto\sqrt{al}$ or
\begin{equation}
 v_0=c_0\sqrt{l(\tan\theta-\mu_r)}, \label{v0}
\end{equation}
where $c_0$ is the proportionality constant.

\begin{figure*}[t]
 \includegraphics[width=512pt]{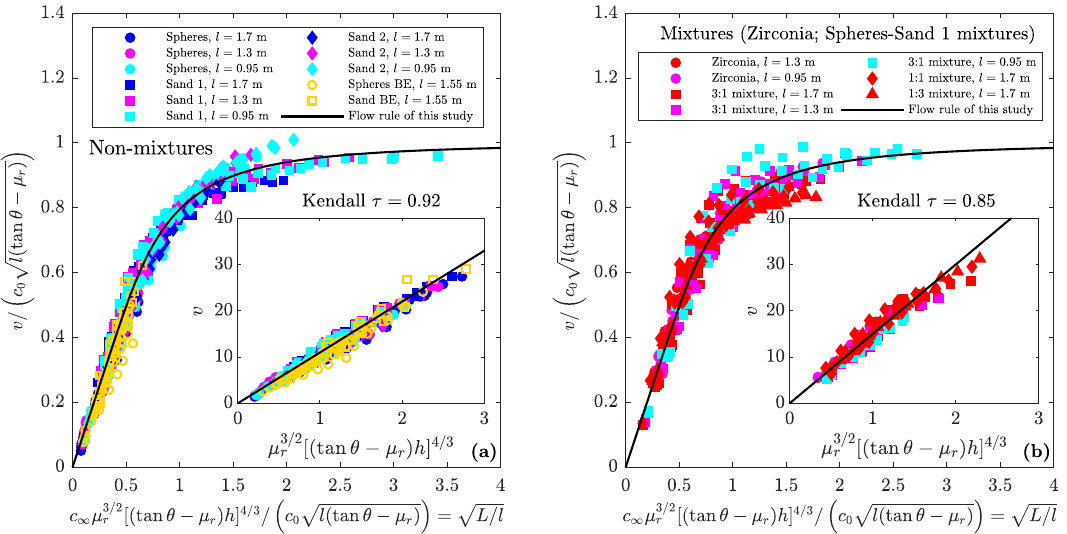}%
 \caption{Test of flow rule unifying undeveloped and fully developed inclined flows against the measurements for (a) nonmixtures and (b) mixtures (Table~\ref{GranularMaterials}). Insets: Data for nearly fully developed conditions ($l>2L$). The solid lines in the main plots and insets correspond to Eqs.~(\ref{NewFlowRule1}), (\ref{NewFlowRule2}) and Eq.~(\ref{vinfty}), respectively, with $c_\infty=11$ and $c_0=1.5$ in (a) and $c_\infty=15$ and $c_0=1.7$ in (b).}
 \label{FlowRule}
\end{figure*}

To model fully developed flows, we borrow a recent insight gained from the unification of aeolian and fluvial sediment transport~\citep{PahtzDuran20}. It was found that the average sediment transport velocity in the steady state is a function of only the excess shear stress applied on the sediment bed surface, $\tau_\mathrm{ex}$, and a flow-strength-independent dimensionless scale that accounts for the environmental conditions, $C_\mathrm{env}$. We assume that an analogous dimensionless relationship, $v_\infty=v_\infty(\tau_\mathrm{ex},C_\mathrm{env})$, also holds for inclined granular flows, in which $\tau_\mathrm{ex}\sim(\tan\theta-\mu_r)h$ (assuming a nearly constant mean volume fraction) and $C_\mathrm{env}\sim\mu_r$ (the main $(h\tan\theta)$-independent dimensionless scale encoding the environmental conditions)~\citep{SuppInclinedFlows}, and propose the simple power-law relationship $v_\infty\propto\mu_r^p[(\tan\theta-\mu_r)h]^q$. We determine the exponents as $p\approx3/2$ and $q\approx4/3$ from weighted-least-squares fitting and maximizing the Kendall rank correlation ($\tau$) coefficients between $v/v_0$ and $v_\infty/v_0$, according to Eq.~(\ref{MergeFunction}), simultaneously for both the nonmixture and mixture measurements~\citep{SuppInclinedFlows}, resulting in two separate rough data collapses (Fig.~\ref{FlowRule}).
The data shown in Fig.~\ref{FlowRule} also include measurements we conducted for some materials at one or two earlier downstream locations, $l=1.3~\mathrm{m}$ and $l=0.95~\mathrm{m}$. We confirmed that $h$ at these locations exhibits approximately the same value as at $l=1.7~\mathrm{m}$, though $v$ is different. In summary, we obtain
\begin{equation}
 v_\infty=c_\infty\mu_r^{3/2}[(\tan\theta-\mu_r)h]^{4/3}, \label{vinfty}
\end{equation}
where $c_\infty$ is a proportionality constant.

To model the function $f$ in Eq.~(\ref{MergeFunction}), we start with the observation that, in previous DEM simulations of inclined flows of spheres, the evolution of $v$ with time $t$ is described by the saturation function $v\approx v_\infty[1-\exp(-t/T)]$~\citep{ParezAharonov15}, implying $v_0\approx v_\infty t/T$ for $t\ll T$. Multiplying with $v_0$, using $l=\int_0^tv_0(t^\prime)\d t^\prime\approx v_0t/2$ and $L\propto v_\infty T$ as $v_\infty$ and $T$ are the only involved saturation scales, it follows $v_0^2\propto v_\infty^2l/L$ for $l\ll L$. Hence, consistent with these relations, we model the spatial evolution of $v$ by the saturation function (other choices are also possible)
\begin{equation}
 v=v_\infty[1-\exp(-l/L)]^{1/2}, \label{NewFlowRule1}
\end{equation}
which implies $L=lv_\infty^2/v_0^2$. Using Eqs.~(\ref{v0}) and (\ref{vinfty}), $L$ is given by
\begin{equation}
 L=\left(c_\infty^2/c_0^2\right)\mu_r^3\left[(\tan\theta-\mu_r)h\right]^{5/3}h. \label{NewFlowRule2}
\end{equation}
Equations~(\ref{NewFlowRule1}) and (\ref{NewFlowRule2}) agree with the measurements, though the values of $c_\infty$ and $c_0$ are different for nonmixtures and mixtures (Fig.~\ref{FlowRule}).

The flow rule derived in this Letter, given by Eqs.~(\ref{NewFlowRule1}) and (\ref{NewFlowRule2}), covers undeveloped and fully developed granular flows down rough inclines. It holds for the dense flow regime, in which air drag can be neglected~\citep{BorzsonyiEcke06}. It provides a first measurements-based estimate for the characteristic distance $L$ a flow needs to fully develop. This may help in estimating the flow energy and destructiveness of natural avalanches. The only previous estimate of $L$ we are aware of is that of Ref.~\citep{ParezAharonov15}, derived using the $\mu(I)$ rheology and supported by DEM simulations of inclined flows of spheres. It predicts $L\propto(\tan\theta-\mu_r)h^3$, which represents a similar dependence on $h$, but a substantially weaker dependence on $(\tan\theta-\mu_r)$ and a much weaker dependence on $\mu_r$ when compared with Eq.~(\ref{NewFlowRule2}). It means the $\mu(I)$ rheology, probably due to its well-documented shortcomings~\citep{Bouzidetal15,Barkeretal15}, is unable to capture the trends of $L$ observed in the measurements.

One may wonder how Eq.~(\ref{NewFlowRule1}) with $L=lv_\infty^2/v_0^2$ would perform if $v_0$ were calculated by Eq.~(\ref{v0}) but $v_\infty$ by an expression different from Eq.~(\ref{vinfty}). In the Supplementary Material~\citep{SuppInclinedFlows}, we test three theoretical predictions for $v_\infty$ using this method --- one from the $\mu(I)$ rheology~\citep{ParezAharonov15}, one from a three-dimensional~\citep{Jenkins07}, and one from a two-dimensional granular kinetic theory~\citep{Jenkins06}, the latter of which very similar to Eq.~(\ref{Scaling2}) --- and allow them to be empirically modified through multiplication with $\mu_r^p$, with an adjustable exponent $p$. Equation~(\ref{vinfty}) quantitatively describes the data best.

It is also noteworthy that our expression for the steady-state velocity $v_\infty$ is simpler than previous ones as it involves only a single parameter representing the granular material, $\mu_r$, which can be interpreted as the dynamic threshold of granular motion. In contrast, expressions involving $h/h_s$, such as Eqs.~(\ref{Scaling1}) and (\ref{Scaling2}), contain two such parameters, $\mu_1$ and $\mu_2$ via Eq.~(\ref{hstop}), and diverge when $\tan\theta$ crosses $\mu_2$. This divergence used to be desirable, since it had been the long-standing consensus that chute flows never cease to accelerate when $\tan\theta\geq\mu_2$, at least in the idealized sidewall-free and air-drag-free case~\citep{Delannayetal07}. However, this consensus has recently been challenged by two-dimensional DEM simulations~\citep{Patroetal21}. They showed that steady, uniform flow states still exist even much above $\mu_2$, though they may no longer be dense and take much longer to be reached. Since $L$ increases and the flow's steady-state mean volume fraction decreases with $\tan\theta$ already before $\mu_2$ is crossed, albeit perhaps more slowly, it is the null hypothesis that no true phase transition occurs at $\mu_2$ until proven otherwise. If this hypothesis holds true, one should not a-priori expect the occurrence of $h_s$ or $\mu_2$ in the scaling of $v_\infty$.

The original intention of using mixtures of equal-sized (but shape-bidisperse) Spheres and Sand~1 was to produce a wider range of granular materials to test our model. This almost works as the model captures the behavior of the mixtures. However, the scaling parameters are clearly different from those of nonmixtures (Fig.~\ref{FlowRule}). We believe this is attributed to segregation observed during the mixture experiments: fine sand grains, well below the median grain size of Sand~1, dominate the near-base layers of the remaining deposit. This will be studied in more detail using numerical simulations in the future.

\begin{acknowledgments}
Z.H. acknowledges financial support from grant National Key R \& D Program of China (2023YFC3008100). Y.W. acknowledges financial support from grant Project Supported by Scientific Research Fund of Zhejiang University (XY2023034). T.P. acknowledges financial support from grants National Natural Science Foundation of China (No.~12350710176, No.~12272344). Z.H. acknowledges financial support from grant National Natural Science Foundation of China (No.~52171276). L.J. acknowledges financial support provided by the Open Research Fund Program of State Key Laboratory of Hydroscience and Engineering (sklhse-2023-B-07).
\end{acknowledgments}

%

\end{document}